\documentclass[onecolumn,aps]{revtex4}
\usepackage{orcidlink}
\usepackage{mathrsfs}
\usepackage{amsfonts}
\usepackage{dcolumn}
\usepackage{bm}
\usepackage{epsfig}
\usepackage{epstopdf} 
\usepackage{pgf,fancyhdr}
\usepackage{braket}
\usepackage{amsmath}
\usepackage{amssymb}
\usepackage{amsthm}
\usepackage{graphicx}
\usepackage{hyperref}
\usepackage{booktabs}
\usepackage{xcolor,amsmath}
\usepackage[title]{appendix}
\hypersetup{
colorlinks=true,
linkcolor=blue,
urlcolor=blue,
citecolor=blue,
pdfborder={0 0 0}
}
\usepackage{calrsfs}
\DeclareMathAlphabet{\pazocal}{OMS}{zplm}{m}{n}
\newtheorem{theorem}{Theorem} 
\newtheorem{corollary}{Corollary}
\theoremstyle{definition} 

\newtheorem{example}{Example}
\theoremstyle{remark}    

\definecolor{RED}{RGB}{255,0,0}
\begin{document}
\title{Enhanced separability criteria based on symmetric measurements}
\author{Yu Lu$^{1}$\orcidlink{0009-0003-0152-2687}}%
\author{Hao-Fan Wang$^{1}$}
\author{Meng Su$^{1}$}%
\author{Zhi-Xi Wang$^{1}$\orcidlink{0000-0002-8341-5142}}
\email[]{wangzhx@cnu.edu.cn}
\author{Shao-Ming Fei$^{1}$\orcidlink{0000-0003-2412-8626}}
\email[]{feishm@cnu.edu.cn}
\affiliation{%
$^{1}$School of Mathematical Sciences, Capital Normal University, Beijing 100048, China}

\begin{abstract}
We present separability criteria based on local symmetric measurements. These experimentally feasible criteria are shown to be more efficient  in detecting entanglement than the existing criteria by detailed examples. Furthermore, we generalize the separability criteria from bipartite to arbitrary multipartite systems. These criteria establish a richer connection between the quantum entanglement and the probabilities of local measurement outcomes.
\end{abstract}

\keywords{Symmetric measurements, Quantum entanglement, Correlation matrix}

\maketitle

\section{Introduction}

Quantum entanglement \cite{Mintert1679022004,Chen0405042005,Breuer0805012006,Vicente0523202007,Zhang0123342007} is the central resource in quantum information processing and quantum computation \cite{nielsen2010quantum}. A primary challenge in the theory of quantum entanglement is the operational detection of entangled states. Consider a $d_X$-dimensional Hilbert space $H_X$ associated with a quantum system $X$. A bipartite state $\rho_{AB}\in H_{A}\otimes H_{B}$ is called separable if it can be written as a convex sum of tensor products of the local states of the subsystems,
\begin{equation}\label{Eq:1}
\rho_{AB}=\sum_{i}p_i\rho^{i}_{A}\otimes \rho^{i}_{B},
\end{equation}
where $p_i\geq 0$ and $\sum_i p_i=1$. Otherwise, $\rho$ is said to be entangled.
A variety of separability criteria have been established to distinguish entangled states from separable ones. Notable examples include the positive partial transposition (PPT) criterion \cite{peres1996separability,horodecki1996necessary,horodecki1997separability}, the realignment criterion \cite{rudolph2003some,horodecki2006separability,chen2002generalized,albeverio2003generalized}, the covariance matrix criterion \cite{guhne2007covariance} and the correlation matrix criterion \cite{de2008further}.

While considerable progress has been made toward developing theoretical tools in entanglement detection, relatively few results \cite{gisin1991bell,yu2003comprehensive,li2010gisin,mingjing2011inequalities} address its experimental implementation for unknown quantum states. In Ref.\cite{durt2010mutually}, the authors introduced the concept of mutually
unbiased bases (MUBs) for two-qubit, many-body and continuous-variable quantum systems, respectively. In Ref.\cite{spengler2012entanglement}, the authors employed MUBs to formulate separability criteria applicable to both bipartite and multipartite systems. In particular, the MUBs-based criterion is necessary and sufficient for isotropic states when the local dimension $d$ is a prime power. A more universally applicable framework is provided by mutually unbiased measurements (MUMs) \cite{kalev2014mutually}, which hold a practical advantage: a complete set of $d+1$ MUMs can be constructed for any $d$-dimensional Hilbert space, regardless of whether $d$ is a prime power. Based on MUMs, a separability criterion applicable to bipartite systems was presented in Ref.\cite{chen2014entanglement}.

The study of SIC-POVMs in the literature typically concentrates on the rank-$1$ case, where all POVM measurement operators are proportional to rank-$1$ projectors. However, the scalability of finding numerical solutions is severely limited with the increasing dimensionality \cite{axioms6030021}. The general SIC-POVMs (GSIC-POVMs) relax the rank-$1$  constraint of the standard SIC-POVMs and are proven to exist in every finite dimension, with explicit constructions given in Ref.\cite{appleby2007symmetric,gour2014construction}. The symmetric $(N,M)$-POVMs \cite{Siudzinska2022} generalize both MUMs and GSIC-POVMs. As an application in Ref.\cite{Lu2025}, the authors derived a separability criterion based on such POVMs. Building on this work, in this paper we develop an enhanced separability criterion which demonstrates a greater capability for detecting entanglement.

The remaining sections of this paper are structured as follows. In Section  \ref{sec:2}, we recall the essential concepts related to \((N,M)\)-POVMs. In Section \ref{sec:3}, we review several entanglement criteria based on symmetric measurements. In Section \ref{sec:4}, we develop separability criteria for bipartite quantum systems by employing \((N,M)\)-POVMs, provide detailed examples to illustrate our results, and extend the criteria from bipartite systems to multipartite cases. A summary and conclusions are provided in Section  \ref{sec:5}.

\section{Preliminaries}\label{sec:2}

The symmetric informationally complete measurements \cite{Siudzinska2022},
$(N, M)$-POVMs, are given by $N$ $d$-dimensional POVMs, $\{E_{\alpha,k} \mid k=1,2,\cdots,M\}$ ($\alpha=1,2,\cdots,N$), each with $M$ measurement operators satisfying
\begin{flalign}\nonumber\label{Eq:2}
\mathrm{tr}(E_{\alpha,k}) &= w, \\ \nonumber
\mathrm{tr}(E_{\alpha,k}^{2}) &= x,\\
\mathrm{tr}(E_{\alpha,k}E_{\alpha,l}) &= y~~(l\neq k),\\
\mathrm{tr}(E_{\alpha,k}E_{\beta,l}) &= z~~(\beta\neq\alpha),\nonumber
\end{flalign}
where
\begin{flalign}\label{Eq:3}
w = \frac{d}{M}, \quad y = \frac{d - Mx}{M(M - 1)}, \quad z = \frac{d}{M^2}
\end{flalign}
and
\begin{flalign}\label{Eq:4}
\dfrac{d}{M^{2}}<x\leq \min\left\{\dfrac{d^{2}}{M^{2}},\dfrac{d}{M}\right\}.
\end{flalign}

In particular, $(N, M)$-POVMs reduce to projective measurements when $ x = \dfrac{d^2}{M^2} $, for which (\ref{Eq:3}) implies that projective $(N, M)$-POVMs are possible only if $M \ge d$. The symmetric measurements attain the informational completeness precisely when $(M - 1)N = d^2 - 1$.

Consider an orthonormal Hermitian operator basis ${G_{0}=I_{d}/\sqrt{d},~G_{\alpha,k}}$ with $\mathrm{tr}(G_{\alpha,k})=0$, where $\alpha=1,\cdots,N$, $k=1,\cdots,M-1$, and $I_d$ is the $d\times d$ identity. The measurement operators of an informationally complete $(N,M)$-POVM are then explicitly given by
\begin{equation}\label{Eq:5}
E_{\alpha,k}=\dfrac{1}{M}I_{d}+tH_{\alpha,k},
\end{equation}
where
\begin{flalign}\label{Eq:6}
 	H_{\alpha,k}=\begin{cases}
 		G_{\alpha}-\sqrt{M}(\sqrt{M}+1)G_{\alpha,k}~~&(k=1,\cdots,M-1),\\[1mm]
 		(\sqrt{M}+1)G_{\alpha}~~&(k=M),
 	\end{cases}
 \end{flalign}
with $G_{\alpha}=\sum\limits_{k=1}^{M-1}G_{\alpha,k}$. The positivity requirement of $E_{\alpha,k}$ imposes the condition,
\begin{equation*}
 	-\dfrac{1}{M}\dfrac{1}{\lambda_{\max}}\leq t \leq \dfrac{1}{M}\dfrac{1}{|\lambda_{\min}|},
 \end{equation*}
where $\lambda_{\text{max}}$ and $\lambda_{\text{min}}$ denote the maximum and minimum eigenvalues, respectively, among all $H_{\alpha,k}$. The parameters $t$ and $x$ are related by the following relation,
\begin{equation}\label{Eq:7}
x=\dfrac{d}{M^{2}}+t^{2}(M-1)(\sqrt{M}+1)^{2}.
\end{equation}

From  (\ref{Eq:2}) and  (\ref{Eq:3}), one obtains the coincidence index corresponding to the probability distribution generated by  $\sum_{\alpha=1}^{N} \sum_{k=1}^{M} E_{\alpha, k}$,
\begin{equation}\label{Eq:8}
C(x, \rho) = \sum_{\alpha=1}^{N} \sum_{k=1}^{M} \langle E_{\alpha, k} \rangle_{\rho}^2 = \frac{d(M^2 x - d) \mathrm{tr}(\rho^2) + d^3 - M^2 x}{dM(M-1)},
\end{equation}
where $\langle E_{\alpha, k} \rangle_{\rho}=\mathrm{tr}(E_{\alpha, k} {\rho})$.
Therefore, the condition $\mathrm{tr}(\rho^2) \leq 1$ implies that $C(x, \rho)$ has the following upper bound,
\begin{equation}\label{Eq:9}
C(x, \rho) \leq \frac{d-1}{d} \frac{d^2 + M^2 x}{M(M-1)}.
\end{equation}

\section{Some entanglement criteria}\label{sec:3}
\subsection{MUMs}

Since MUBs cannot guarantee complete existence in non-prime-power dimensions, Chen et al.\cite{chen2014entanglement} introduced MUMs into the field of entanglement detection in 2014. MUMs do not require measurement operators to be rank-$1$ projections and can construct complete sets of \(d+1\) measurements in any dimension, thus providing a more universal tool for entanglement detection. Chen et al. proposed a separability criterion by using MUMs
\begin{equation}\label{Eq:10}
J(\rho_{AB}) = \sum_{b=1}^{d+1}\sum_{n=1}^{d} \mathrm{tr}\bigl(P_n^{(b)} \otimes Q_n^{(b)} \rho_{AB}\bigr) \leq 1 + \kappa,
\end{equation}
where \(\{P^{(b)}\}_{b=1}^{d+1}\) and \(\{Q^{(b)}\}_{b=1}^{d+1}\) are any two sets  of \( d+1 \) MUMs on $ \mathbb{C}^d $ with the same parameter $ \kappa $,
$\mathcal{P}^{(b)} = \{P_n^{(b)}\}_{n=1}^d, \quad \mathcal{Q}^{(b)} = \{Q_n^{(b)}\}_{n=1}^d, \quad b = 1, 2, \dots, d+1,$ $J(\rho_{AB}) = \sum_{b=1}^{d+1} \sum_{n=1}^d \mathrm{tr}(P_n^{(b)} \otimes Q_n^{(b)} \rho_{AB}).$

Subsequently, in 2015, Liu et al. \cite{liu2015separability} extended the MUMs criterion to multipartite qudit systems, allowing for subsystems of different dimensions (\(d_1 \neq d_2\)) and enabling the detection of \(k\)-separability for the first time, thereby broadening its applicability. In the same year, Shen et al. \cite{shen2015entanglement} introduced a decomposition structure of \(\rho_{AB} - \rho^A \otimes \rho^B\) and proposed a tighter form of the criterion:
\begin{align}\label{Eq:11}
   S(\rho_{AB})&  = \sum_{b=1}^{d+1} \sum_{n=1}^{d} \left| \mathrm{tr}\left( P_n^{(b)} \otimes Q_n^{(b)} (\rho_{AB} - \rho^A \otimes \rho^B) \right) \right| \nonumber\\
 &\leq \sqrt{ \left( 1 + \kappa - \sum_{b=1}^{d+1} \sum_{n=1}^{d} \bigl(\mathrm{tr}(P_n^{(b)} \rho^A) \bigr)^2 \right) \left( 1 + \kappa - \sum_{b=1}^{d+1} \sum_{n=1}^{d} \bigl( \mathrm{tr}(Q_n^{(b)} \rho^B) \bigr)^2 \right) }.
\end{align}

After 2017, Liu et al.\cite{liu2017separability} and Lu et al.\cite{lu2018new} further generalized the MUMs criterion to more general multipartite systems and introduced free parameters to optimize detection performance.

\subsection{GSIC-POVMs}
Since the introduction of GSIC-POVMs by Gour and Kalev in 2014, this class of measurements has rapidly become an important tool for entanglement detection due to its constructibility in arbitrary dimensions and the absence of the requirement for measurement operators to be rank-$1$ projections. Compared to MUMs, GSIC-POVMs require only \(d^2\) joint local measurements, significantly reducing the complexity of experimental implementation. In 2015, Chen et al.\cite{chen2015general} applied GSIC-POVMs to entanglement detection for the first time and proposed a fundamental criterion:
\begin{equation}\label{Eq:12}
J_a(\rho_{AB}) = \sum_{j=1}^{d^2} \mathrm{tr}(P_j \otimes Q_{j}\rho_{AB}) \leq \frac{ad^2 + 1}{d(d+1)},
\end{equation}
where \(\{P_j\}\) and \(\{Q_j\}\) are any two sets  of general symmetric informationally complete measurements on \( \mathbb{C}^d \) with the same parameter \( a \), \( J_a(\rho_{AB}) = \sum_{j=1}^d \mathrm{tr}(P_j \bigotimes Q_j \rho_{AB}) \).

During the subsequent phase of improvement and extension $(2016 - 2018)$, Xi et al.\cite{xi2016entanglement} generalized the GSIC-POVMs criterion to multipartite systems, while Shen et al.\cite{shen2018improved} introduced the decomposition structure of $\rho_{AB} - \rho^A \otimes \rho^B$ 
and proposed a tighter criterion based on the trace norm:
\begin{equation}\label{Eq:13}
\|M^{(P,Q)}(\rho_{AB})\|_{\mathrm{tr}} \leq \sqrt{\frac{ad^2 + 1}{d(d+1)} - \sum \mathrm{tr}(P_i\rho^A)^2} \sqrt{\frac{ad^2 + 1}{d(d+1)} - \sum \mathrm{tr}(Q_i\rho^B)^2},
\end{equation}
which performs better in detecting Werner states and bound entangled states, significantly improving detection efficiency.

Entering the phase of generalization to high-dimensional and multipartite systems $(2018 - 2022)$, Lai et al.\cite{lai2018entanglement} further extended the GSIC-POVMs criterion to bipartite systems of arbitrary dimensions and demonstrated its superiority on \(3 \times 3\) bound entangled states. Li and Chen \cite{Li_2022}, in 2022, generalized the criterion to tripartite and multipartite systems, proposing a criterion based on the correlation matrix:
\begin{equation}\label{Eq:14}
\|G^{ABC}\|_{\mathrm{tr}} \leq \prod_{i=1}^3 \sqrt{\frac{a_i d_i^2 + 1}{d_i (d_i + 1)}},
\end{equation}
and verified its effectiveness through typical examples such as GHZ states. This series of works not only deepened the application of GSIC-POVMs in entanglement detection but also provided more efficient theoretical tools for experimental detection of multipartite quantum systems.

\subsection{$(N,M)$-POVMs}

In 2022, Siudzi\'{n}ska introduced the concept of \((N, M)\)-POVMs in Ref.\cite{Siudzinska2022}, which provides a unified framework encompassing MUMs and GSIC-POVMs as special cases. These measurements satisfy the relation
\[
(M - 1)N = d^{2} - 1,
\]
and include four typical configurations:
\begin{itemize}
    \item \(M = d^{2},\ N = 1\): GSIC-POVMs;
    \item \(M = d,\ N = d + 1\): MUMs;
    \item \(M = 2,\ N = d^{2} - 1\);
    \item \(M = d + 2,\ N = d - 1\).
\end{itemize}
This unified construction offers a versatile theoretical foundation for entanglement detection.

Based on the \((N, M)\)-POVM framework, in 2023, Tang \cite{tang2023enhancing} proposed an entanglement criterion applicable to bipartite systems of arbitrary dimensions. The author considered bipartite states  $\rho_{AB}\in H_A \otimes H_B$, and two sets of $(N, M)$-POVMs:
$
\{ E_{\alpha,k} | \alpha = 1, 2, \cdots, N_A;\, k = 1, 2, \cdots, M_A \}
$
with the efficiency parameter $x_A$ and
$
\{ E_{\beta,l} | \beta = 1, 2, \cdots, N_B;\, l = 1, 2, \cdots, M_B \}
$
with the efficiency parameter $x_B$ for the two subsystems $H_A$ and $H_B$ with dimensions $d_A$ and  $d_B$, respectively.  Let $\mathcal{P}$ denote the matrix with entries given by $\langle E_{\alpha,k} \otimes E_{\beta,l} \rangle = \mathrm{tr}(E_{\alpha,k} \otimes E_{\beta,l} \rho_{AB})$. If the quantum state \(\rho_{AB}\) is separable in bipartite systems, one has
\begin{equation}\label{Eq:15}
\|\mathcal{P}\|_{\text{tr}} \leq 
\sqrt{\frac{d_A - 1}{d_A}\cdot\frac{d_A^{2} + M_A^{2}x_A}{M_A(M_A-1)}}\;
\sqrt{\frac{d_B - 1}{d_B}\cdot\frac{d_B^{2} + M_B^{2}x_B}{M_B(M_B-1)}},
\end{equation}
where \(\|\mathcal{P}\|_{\text{tr}}\) denotes the trace norm of the correlation matrix.   This criterion exhibits superior performance in detecting entanglement of isotropic states and Bell diagonal states.

Subsequently, Tang and Wu \cite{tang2023enhancing} extended the \((N, M)\) - POVMs based criterion to tripartite and multipartite systems. For a tripartite state \(\rho_{ABC}\) with subsystems of dimensions \(d_1, d_2, d_3\) and corresponding measurement parameters \(x_1, x_2, x_3\), they introduced a correlation-matrix-based criterion:
\begin{equation}\label{Eq:16}
\max \left\{ \| \mathcal{P}^{\text{ABC}} \|_{\text{tr}}, \| \mathcal{P}^{\text{BAC}} \|_{\text{tr}}, \| \mathcal{P}^{\text{CAB}} \|_{\text{tr}} \right\}
 \leq 
\prod_{i=1}^{3} \sqrt{\frac{d_i - 1}{d_i}\cdot\frac{d_i^{2} + M_i^{2}x_i}{M_i(M_i-1)}}.
\end{equation}
This formulation not only unifies the existing MUMs and GSIC-POVMs criteria but also demonstrates enhanced versatility and detection power in many-body quantum systems.

In 2025, Lu et al.\cite{Lu2025}  constructed the matrix $\mathcal{M}_{\mu,\nu}^{l}(\rho_{AB})$ for a bipartite state $\rho_{AB}$,
\begin{align}\label{Eq:17}
\mathcal{M}_{\mu,\nu}^{l}(\rho_{AB})=\begin{pmatrix}
	\mu\nu J_{l\times l}& \mu \omega_{l}(\sigma)^T\\[1mm]
	\nu \omega_{l}(\tau)& \mathcal{P}(\rho_{AB})
\end{pmatrix},
\end{align}
where
\begin{align}\label{Eq:18}
\tau = \left( \begin{array}{c}
\mathrm{tr}(E_{1,1} \rho_A) \\
\mathrm{tr}(E_{1,2} \rho_A) \\
\vdots \\
\mathrm{tr}(E_{1,M_A} \rho_A) \\
\mathrm{tr}(E_{2,1} \rho_A) \\
\mathrm{tr}(E_{2,2} \rho_A) \\
\vdots \\
\mathrm{tr}(E_{2,M_A} \rho_A) \\
\vdots \\
\mathrm{tr}(E_{N_A,M_A} \rho_A)
\end{array} \right), \qquad
\sigma = \left( \begin{array}{c}
\mathrm{tr}(E_{1,1} \rho_B) \\
\mathrm{tr}(E_{1,2} \rho_B) \\
\vdots \\
\mathrm{tr}(E_{1,M_B} \rho_B) \\
\mathrm{tr}(E_{2,1} \rho_B) \\
\mathrm{tr}(E_{2,2} \rho_B) \\
\vdots \\
\mathrm{tr}(E_{2,M_B} \rho_B) \\
\vdots \\
\mathrm{tr}(E_{N_B,M_B} \rho_B)
\end{array} \right),
\end{align}
with $\rho_{A}$ ($\rho_{B}$) being the reduced density matrix obtained by tracing over the subsystem $H_A$ ($H_B$) of $\rho_{AB}$, $\mu$ and $\nu$ are real numbers, $l$ is a natural number, $J_{l \times l}$ is the matrix with all $l \times l$ entries being $1$, $\omega_{l}(X)=\underbrace{(X, \cdots, X)}_{l}$, and $T$ denotes the transpose of a matrix. They showed that if $\rho_{AB}\in\mathcal{H}_{A}\otimes\mathcal{H}_{B}$ is separable, then
\begin{eqnarray}\label{Eq:19}
\left\|\mathcal{M}_{\mu, \nu}^l \left(\rho_{AB}\right)\right\|_{\mathrm{tr}} \leq \sqrt{\left(l \mu^{2}+\frac{(d_{A}-1)(d_{A}^2+M_{A}^2x_A)}{d_{A}M_{A}(M_{A}-1)}\right)\left(l \nu^{2}+\frac{(d_{B}-1)(d_{B}^2+M_{B}^2x_B)}{d_{B}M_{B}(M_{B}-1)}\right)}.
\end{eqnarray}

\section{Main Results}\label{sec:4}

After reviewing the progress of research on entanglement criteria based on (N,M)-POVMs, we note that in Ref.\cite{Lu2025}, the detection efficiency for bipartite systems was improved by constructing a specific matrix form. However, this framework can still be further generalized and optimized under a more general parameterized structure. Motivated by this, we propose a more flexible form of the criterion in this paper. By introducing adjustable parameter vectors \(a\) and \(b\), we construct an extended correlation matrix \(\mathcal{Q}_{a,b}(\rho_{AB})\), which not only preserves the advantages of the original method but also enhances its capability to detect different types of entangled states. In the following, we will elaborate on the construction of this new criterion and its theoretical basis.


\begin{theorem}\label{th:1}
If $\rho_{AB}$ is a bipartite separable state, then
\begin{equation}\label{Eq:20}
	\|\mathcal{Q}_{a, b}(\rho_{AB})\|_{\rm tr}\leq\sqrt{|a|^{2}+\dfrac{(d_{A}-1)(M_{A}^{2}x_{A}
+d_{A}^{2})}{d_{A}M_{A}(M_{A}-1)}}\sqrt{|b|^{2}
+\dfrac{(d_{B}-1)(M_{B}^{2}x_{B}+d_{B}^{2})}{d_{B}M_{B}(M_{B}-1)}},
\end{equation}
where  
\begin{align}\label{Eq:21}
\mathcal{Q}_{a, b}(\rho_{AB})=\begin{pmatrix}
	ab^{T} & a \sigma^{T}\\
	\tau b^{T} & \mathcal{P}(\rho_{AB})
\end{pmatrix},
\end{align}
$a\in\mathbb{R}^{m}$, $b\in\mathbb{R}^{n}$, $\tau$ and $\sigma$  are defined  in Eq.(\ref{Eq:18}), and $T$ denotes the transpose of a matrix. $\|X\|_{\mathrm{tr}}=\mathrm{tr} \sqrt{X^{\dagger}X}$ denotes the trace norm of $X$.
\end{theorem}

\begin{proof}
If $\rho_{AB} $ is separable, it has the form of (\ref{Eq:1}). By using (\ref{Eq:9}), we have
 \begin{align}\label{Eq:22}
\|\mathcal{Q}_{a,b}\left(\rho_{AB}\right)\|_{\mathrm{tr}}&=\|\sum\limits_{i}p_{i}\mathcal{Q}_{a,b}\left(\rho_{i}^{A}\otimes \rho_{i}^{B}\right)\|_{\mathrm{tr}}\nonumber\\
&\leq \sum\limits_{i}p_{i}\|\mathcal{Q}_{a,b}\left(\rho_{i}^{A}\otimes \rho_{i}^{B}\right)\|_{\mathrm{tr}}\nonumber\\
&=\left\|\left(\begin{array}{cc}ab ^{T} & a \sigma_{i_B}^T \\ \tau_{i_A} b^T  & \tau_{i_A}\sigma_{i_B} ^\mathrm{T}\end{array}\right)\right\|_{\mathrm{tr}} \nonumber\\
&=\left\|\left(\begin{array}{cc}a & \nonumber\\ \tau_{i_A} & \end{array}\right)\left(\begin{array}{cc}b^T & \sigma_{i_B}^\mathrm{T}\end{array}\right)\right\|_{\mathrm{tr}} \nonumber\\
&=\left\|\left(\begin{array}{cc}a & \nonumber\\ \tau_{i_A} & \end{array}\right)\right\|_{\mathrm{tr}}\left\|\left(\begin{array}{cc}b^{T} & \sigma_{i_B}^\mathrm{T}\end{array}\right)\right\|_{\mathrm{tr}} \nonumber\\
&\leq \sqrt{|a|^{2}+\dfrac{(d_{A}-1)(M_{A}^{2}x_{A}+d_{A}^{2})}{d_{A}M_{A}(M_{A}-1)}}
\sqrt{|b|^{2}+\dfrac{(d_{B}-1)(M_{B}^{2}x_{B}+d_{B}^{2})}{d_{B}M_{B}(M_{B}-1)}}.
\end{align}
The first equality follows from
 \begin{align}\label{Eq:23}
\mathcal{Q}_{a,b}\left(\sum\limits_{i}p_{i}\rho_{i}^{A}\otimes \rho_{i}^{B}\right)&=\begin{pmatrix}
	ab^{T} & a \sigma_B^{T}\nonumber\\
	\tau_A b^{T} & \mathcal{P}(\sum\limits_{i}p_{i}\rho_{i}^{A}\otimes \rho_{i}^{B})\end{pmatrix}\nonumber\\
&=\begin{pmatrix}
	\sum\limits_{i}p_{i}ab^{T} & \sum\limits_{i}p_{i}a \sigma_{i_B}^{T}\nonumber\\
	\sum\limits_{i}p_{i}\tau_{i_A} b^{T} & \sum\limits_{i}p_{i}\mathcal{P}(\rho_{i}^{A}\otimes \rho_{i}^{B})\end{pmatrix}\nonumber\\
&=\sum\limits_{i}p_{i}\begin{pmatrix}
	ab^{T} & a \sigma_{i_B}^{T}\nonumber\\
	\tau_{i_A} b^{T} & \mathcal{P}(\rho_{i}^{A}\otimes \rho_{i}^{B})\end{pmatrix}\nonumber\\
&=\sum\limits_{i}p_{i}\mathcal{Q}_{a,b}\left(\rho_{i}^{A}\otimes \rho_{i}^{B}\right),
\end{align}
where
\begin{align*}
\tau_A = \left( \begin{array}{c}
\mathrm{tr}(E_{1,1} \sum\limits_{i}p_{i}\rho_{i}^{A}) \\
\mathrm{tr}(E_{1,2} \sum\limits_{i}p_{i}\rho_{i}^{A}) \\
\vdots \\
\mathrm{tr}(E_{N_A,M_A} \sum\limits_{i}p_{i}\rho_{i}^{A})
\end{array} \right), \qquad
\sigma_B = \left( \begin{array}{c}
\mathrm{tr}(E_{1,1} \sum\limits_{i}p_{i}\rho_{i}^{B}) \\
\mathrm{tr}(E_{1,2} \sum\limits_{i}p_{i}\rho_{i}^{B}) \\
\vdots \\
\mathrm{tr}(E_{N_B,M_B} \sum\limits_{i}p_{i}\rho_{i}^{B})
\end{array} \right),
\end{align*}
\begin{align*}
\tau_{i_A} = \left( \begin{array}{c}
\mathrm{tr}(E_{1,1} \rho_{i}^{A}) \\
\mathrm{tr}(E_{1,2} \rho_{i}^{A}) \\
\vdots \\
\mathrm{tr}(E_{N_A,M_A} \rho_{i}^{A})
\end{array} \right), \qquad
\sigma_{i_B} = \left( \begin{array}{c}
\mathrm{tr}(E_{1,1} \rho_{i}^{B}) \\
\mathrm{tr}(E_{1,2} \rho_{i}^{B}) \\
\vdots \\
\mathrm{tr}(E_{N_B,M_B} \rho_{i}^{B})
\end{array} \right),
\end{align*}
The last equality follows from 
$
\|AB\|_{\mathrm{tr}} \leq \|A\|_{\mathrm{tr}} \|B\|_{\mathrm{tr}}
$
for any matrices $A$ and $B$, with the equality holding if and only if \(A\) and \(B\) are both rank-$1$ matrices. Thus, Eq.(\ref{Eq:20}) is proved.
\end{proof}

In particular, taking $(N,M)$-POVMs to be GSIC-POVMs and MUMs, we have the following corollaries:

\begin{corollary}\label{co:1}
Consider two GSIC-POVMs $\{M^A_{\alpha}\}_{\alpha=1}^{d_A^{2}}$ with parameter $a_A$ and $\{M^B_{\beta}\}_{\beta=1}^{d_B^{2}}$ with parameter $a_B$.
If a state $\rho_{AB}\in H_{A}\otimes H_{B}$ is separable, then
\begin{eqnarray}\label{Eq:24}
\left\|\mathcal{Q}_{a, b} \left(\rho_{AB}\right)\right\|_{\mathrm{tr}} \leq \sqrt{\left(|a|^{2}+\frac{a_{A}d_{A}^{2}+1}{d_{A}(d_{A}+1)}\right)\left(|b|^{2}+\frac{a_{B}d_{B}^{2}+1}{d_{B}(d_{B}+1)}\right)},
\end{eqnarray}
where \begin{align}\label{Eq:25}
\mathcal{Q}_{a,b}(\rho_{AB})=\begin{pmatrix}
	a b^T& a \zeta^T\\[1mm]
	\varsigma b^T & \mathcal{G}(\rho_{AB})
\end{pmatrix},
\end{align}
with
\begin{eqnarray*}
\varsigma = \left( \begin{array}{c}
\mathrm{tr}(M_1^A \rho_A) \\
\mathrm{tr}(M_2^A \rho_A) \\
\vdots \\
\mathrm{tr}(M_{d^2_A}^A \rho_A)
\end{array} \right), \qquad
\zeta = \left( \begin{array}{c}
\mathrm{tr}(M_1^B \rho_B) \\
\mathrm{tr}(M_2^B \rho_B) \\
\vdots \\
\mathrm{tr}(M_{d^2_B}^B \rho_B)
\end{array} \right),
\end{eqnarray*}
$\mathcal{G}(\rho_{AB})$ is a matrix with entries given by $\mathrm{tr}(M^A_{\alpha}\otimes M^B_{\beta}\rho_{AB})$.
\end{corollary}

\begin{corollary}\label{co:2}
Let $P^{(b_A)} = \{P_{n_A}^{(b_A)}\}_{{n_A}=1}^{d_A}$ with the parameter $\kappa_A$ and $Q^{(b_B)} = \{Q_{{n_B}}^{(b_B)}\}_{{n_B}=1}^{d_B}$ with the parameter $\kappa_B$ be the MUMs on the subsystems $H_A$ and $H_B$, respectively. Denote by $\mathcal{J}(\rho_{AB})$ the matrix with entries given by the probabilities $\mathrm{tr}[(P_{n_A}^{(b_A)} \otimes Q_{n_B}^{(b_B)})\rho_{AB}]$, where $n_A, b_A$ and $n_B, b_B$ are the row and the column indices, respectively. For any separable state $\rho_{AB}$ in $H_{A} \otimes H_B$, we have

\begin{eqnarray}\label{Eq:26}
\left\|\mathcal{Q}_{a, b} \left(\rho_{AB}\right)\right\|_{\mathrm{tr}} \leq \sqrt{\left(|a|^{2} + 1 + \kappa_A\right)\left(|b|^{2}+ 1 + \kappa_B\right)},
\end{eqnarray}

where \begin{align}\label{Eq:27}
\mathcal{Q}_{a,b}(\rho_{AB})=\begin{pmatrix}
	a b^T& a \eta^T\\[1mm]
	\theta b^T & \mathcal{J}(\rho_{AB})
\end{pmatrix},
\end{align}
\begin{eqnarray*}
\theta = \left( \begin{array}{c}
\mathrm{tr}(P_1^{(1)} \rho_A) \\
\mathrm{tr}(P_2^{(1)} \rho_A) \\
\vdots \\
\mathrm{tr}(P_{n_A}^{(1)} \rho_A)\\
\mathrm{tr}(P_1^{(2)} \rho_A) \\
\mathrm{tr}(P_2^{(2)} \rho_A) \\
\vdots \\
\mathrm{tr}(P_{n_A}^{(2)} \rho_A)\\
\mathrm{tr}(P_{n_A}^{(3)} \rho_A)\\
\vdots \\
\mathrm{tr}(P_{n_A}^{(b_A)} \rho_A)
\end{array} \right), \qquad
\eta = \left( \begin{array}{c}
\mathrm{tr}(Q_1^{(1)} \rho_A) \\
\mathrm{tr}(Q_2^{(1)} \rho_A) \\
\vdots \\
\mathrm{tr}(Q_{n_B}^{(1)} \rho_B)\\
\mathrm{tr}(Q_1^{(2)} \rho_B) \\
\mathrm{tr}(Q_2^{(2)} \rho_B) \\
\vdots \\
\mathrm{tr}(Q_{n_B}^{(2)} \rho_B)\\
\mathrm{tr}(Q_{n_B}^{(3)} \rho_B)\\
\vdots \\
\mathrm{tr}(Q_{n_B}^{(b_B)} \rho_B)
\end{array} \right).
\end{eqnarray*}
\end{corollary}

{\it [Remark 1]} When both parameters $a$ and $b$ are zero vectors, Theorem \ref{th:1} reduces to the conclusion presented in Ref.\cite{tang2023enhancing}, namely, if $\rho_{AB}$ is separable then
\begin{eqnarray}\label{Eq:28}
\|\mathcal{P}(\rho_{AB})\|_{\mathrm{tr}}\leq  \sqrt{\left(\frac{(d_{A}-1)(d_{A}^2+M_{A}^2x_A)}{d_{A}M_{A}(M_{A}-1)}\right)
\left(\frac{(d_{B}-1)(d_{B}^2+M_{B}^2x_B)}{d_{B}M_{B}(M_{B}-1)}\right)}.
\end{eqnarray}

{\it [Remark 2]} If one takes $a$ and $b$ to be zero vectors, Corollary \ref{co:1} reduces to GSIC-POVMs criterion. With \( a_A \) and \( a_B \) being the efficiency parameters for subsystems $A$ and $B$ respectively, if \( rho_{AB} \) is separable, then \cite{lai2018entanglement}
\begin{equation}\label{Eq:29}
\|\mathcal{G}(\rho_{AB})\|_{\mathrm{tr}} \leq \sqrt{\frac{a_A d_A^2 + 1}{d_A (d_A + 1)}} \sqrt{\frac{a_B d_B^2 + 1}{d_B (d_B + 1)}}.
\end{equation}

{\it [Remark 3]} When $a$ and $b$ are zero vectors, as a consequence of Corollary \ref{co:2}, a criterion based on MUMs is obtained, i.e., if $\rho_{AB} $ is separable,
\begin{equation}\label{Eq:30}
\|\mathcal{J}(\rho_{AB})\|_{\mathrm{tr}} \leq \sqrt{\kappa_A + 1} \sqrt{\kappa_B + 1},
\end{equation}
where $\kappa_A,\,\kappa_B$ are the efficiency parameters.

{\it [Remark 4]}  The matrix construction in Theorem \ref{th:1} can also be applied to generalized equiangular measurements (GEAMs). According to Ref.\cite{siudzinska2024informationally}, a GEAM can be constructed from generalized symmetric measurements by
\[
P_{\alpha,k}=\gamma_\alpha E_{\alpha,k},
\]
where \(\gamma_\alpha\) is a probability distribution. The corresponding parameters are given by
\[
a_\alpha=\gamma_\alpha w_\alpha,\quad 
b_\alpha=\frac{x_\alpha}{w_\alpha^2},\quad 
c_\alpha=\frac{y_\alpha}{w_\alpha^2},\quad 
f=\frac{z_{\alpha\beta}}{w_\alpha w_\beta}.
\]
This shows that the parameter \(b_\alpha\) in GEAMs plays a role analogous to the parameter \(x\) in \((N,M)\)-POVMs, and 
both types of measurements can be constructed using Hermitian operator bases.

For the subclass of GEAMs that are conical two-designs and satisfy
\[
a_\alpha^2(b_\alpha-c_\alpha)=S,
\]
the index of coincidence is
\[
C=S\left(\mathrm{tr}\rho_{AB}^2-\frac{1}{d}\right)+\varkappa,
\]
where \(\varkappa=(1/d)\sum_{\alpha=1}^N a_\alpha\gamma_\alpha\). Hence
\[
C\leq C^{\max}:=\frac{d-1}{d}S+\varkappa.
\]
Replacing the coincidence-index bound of \((N,M)\)-POVMs by this GEAM bound, one obtains, for any separable bipartite state \(\rho_{AB}\),
\[
\left\|Q_{\mathbf u,\mathbf v}^{\rm GEAM}(\rho_{AB})\right\|_{\rm tr}
\leq
\sqrt{\|\mathbf u\|^2+C_A^{\max}}\,
\sqrt{\|\mathbf v\|^2+C_B^{\max}} .
\]
This shows that the present method is not limited to \((N,M)\)-POVMs. For example, choosing distinct parameters \(b_\alpha\) with \(\gamma_\alpha=1/N\) gives GEAMs outside the \((N,M)\)-POVM framework, while the same matrix construction remains applicable.

As a simple illustration beyond the \((N,M)\)-POVM framework, one may take \(d=2\), \(N=2\), \(M_1=2\), \(M_2=3\), and choose
\[
\gamma_1=\gamma_2=\frac12,\qquad 
b_1=\frac{7}{12},\qquad 
b_2=\frac34 .
\]
Then
\[
a_1=\frac{d\gamma_1}{M_1}=\frac12,\qquad 
a_2=\frac{d\gamma_2}{M_2}=\frac13,
\]
and
\[
c_\alpha=\frac{M_\alpha-db_\alpha}{d(M_\alpha-1)} .
\]
A direct calculation gives
\[
c_1=\frac{5}{12},\qquad c_2=\frac38,
\]
so that
\[
a_1^2(b_1-c_1)=a_2^2(b_2-c_2)=\frac1{24}.
\]
Thus this GEAM satisfies \(a_\alpha^2(b_\alpha-c_\alpha)=S\) with \(S=1/24\), and hence belongs to the conical two-design subclass. Since \(M_1\neq M_2\) and \(b_1\neq b_2\), this example is not an \((N,M)\)-POVM, while the same extended matrix construction remains applicable. This observation shows that our approach is not restricted to \((N,M)\)-POVMs and may be used to derive enhanced separability criteria for more general informationally overcomplete measurements.

Having discussed this possible extension to GEAMs, we now return to the \((N,M)\)-POVMs framework and clarify the relation between our matrix \(Q_{a,b}(\rho_{AB})\) and the matrix \(M^l_{\mu,\nu}(\rho_{AB})\) introduced in Ref.\cite{Lu2025}.
 
A comparison of Eqs.(\ref{Eq:21}) and (\ref{Eq:17}) shows that if we choose \[
a = (\mu, \mu, \dots, \mu)^T \in \mathbb{R}^l, \quad b = (\nu, \nu, \dots, \nu)^T \in \mathbb{R}^l,
\]
then \(ab^T = \mu\nu J_{l\times l}\), and \(a\sigma^T = \mu \omega_l(\sigma)^T\), \(\tau b^T = \nu \omega_l(\tau)\). Hence,
\(
\mathcal{Q}_{a,b}(\rho_{AB}) = \mathcal{M}_{\mu,\nu}^l(\rho_{AB}).
\)

Next, we illustrate our results by several examples.

\begin{example}\label{Ex:1}
We consider the following state by mixing $\rho_{AB}$ with white noise,
\begin{equation}\label{Eq:31}
\rho_{p}=\dfrac{1-p}{9}I_{9}+p\rho_{AB},
\end{equation}
where
\begin{equation}\label{Eq:32}
\rho_{AB} = \dfrac{1}{4}\left(I_9-\sum\limits_{i=0}^{4}\ket{\varphi_{i}}\bra{\varphi_{i}}\right)
\end{equation}
is a $3\times 3$ PPT entangled state with
   \begin{align*}
    &\ket{\varphi_{0}}=\ket{0}\otimes\dfrac{1}{\sqrt{2}}(\ket{0}-\ket{1}), ~~ \ket{\varphi_{1}}=\dfrac{1}{\sqrt{2}}(\ket{0}-\ket{1})\otimes\ket{2}, \\ &\ket{\varphi_{2}}=\ket{2}\otimes\dfrac{1}{\sqrt{2}}(\ket{1}-\ket{2}),~~  \ket{\varphi_{3}}=\dfrac{1}{\sqrt{2}}(\ket{1}-\ket{2})\otimes\ket{0},\\ &\ket{\varphi_{4}}=\dfrac{1}{\sqrt{3}}(\ket{0}+\ket{1}+\ket{2})\otimes\dfrac{1}{\sqrt{3}}(\ket{0}+\ket{1}+\ket{2}).
    \end{align*}

Consider the $(8,2)$-POVMs with its Hermitian basis operators $G_{\alpha, k}$ defined via the Gell-Mann matrices \cite{gell1962symmetries},
\begin{align*}
  &G_{11}=\dfrac{1}{\sqrt{2}}\begin{pmatrix}
    	0 & 1 & 0\\
    	1 & 0 & 0\\
    	0 & 0 & 0
    \end{pmatrix},~
    G_{21}=\dfrac{1}{\sqrt{2}}\begin{pmatrix}
    	0 & -\mathrm{i} & 0\\
    	\mathrm{i} & 0 & 0\\
    	0 & 0 & 0
    \end{pmatrix},~
    G_{31}=\dfrac{1}{\sqrt{2}}\begin{pmatrix}
    	0 & 0 & 1\\
    	0 & 0 & 0\\
    	1 & 0 & 0
    \end{pmatrix},~
    G_{41}=\dfrac{1}{\sqrt{2}}\begin{pmatrix}
    	0 & 0 & -\mathrm{i}\\
    	0 & 0 & 0\\
    	\mathrm{i} & 0 & 0
    \end{pmatrix}, \\[1mm]
  & G_{51}=\dfrac{1}{\sqrt{2}}\begin{pmatrix}
    	0 & 0 & 0\\
    	0 & 0 & 1\\
    	0 & 1 & 0
    \end{pmatrix},~
    G_{61}=\dfrac{1}{\sqrt{2}}\begin{pmatrix}
    	0 & 0 & 0\\
    	0 & 0 & -\mathrm{i}\\
    	0 & \mathrm{i} & 0
    \end{pmatrix},~
  G_{71}=\dfrac{1}{\sqrt{2}}\begin{pmatrix}
    	1 & 0 & 0\\
    	0 & -1 & 0\\
    	0 & 0 & 0
    \end{pmatrix},~
    G_{81}=\dfrac{1}{\sqrt{6}}\begin{pmatrix}
    	1 & 0 & 0\\
    	0 & 1 & 0\\
    	0 & 0 & -2
    \end{pmatrix}.
\end{align*}
Here, the parameter $x=\frac{3}{4}+t^{2}(\sqrt{2}+1)^{2}$ with $t\in[-0.2536,0.2536]$.
In particular, for $a = (0.1, 0.1)^T$, $b =(0.05, 0.07)^T$, and $t = 0.01$, using Theorem \ref{th:1} we obtain
\begin{align}\label{Eq:33}
f_1(p) \nonumber &\equiv \left\|\mathcal{Q}_{a, b} \left(\rho_{p}\right)\right\|_{\mathrm{tr}} - \sqrt{\left(|a|^2+\frac{(d_{A}-1)(d_{A}^2+M_{A}^2x_A)}{d_{A}M_{A}(M_{A}-1)}\right)\left(|b|^2+\frac{(d_{B}-1)(d_{B}^2+M_{B}^2x_B)}{d_{B}M_{B}(M_{B}-1)}\right)}\\  &\approx 0.00089177p -0.0007867,
\end{align}
which is  positive (i.e., $\rho_p$ is entangled) when $ 0.8821786 \leq p \leq 1$.
 From Theorem $1$ in Ref.\cite{Lu2025}, for $\mu = 0.1$, $\upsilon =0.05$, $l = 2$ and $t = 0.01$,  one has
\begin{align}\label{Eq:34}
g_1(p) \nonumber &\equiv \left\|\mathcal{Q}_{\mu, \nu}^l \left(\rho_{p}\right)\right\|_{\mathrm{tr}} - \sqrt{\left(l \mu^{2}+\frac{(d_{A}-1)(d_{A}^2+M_{A}^2x_A)}{d_{A}M_{A}(M_{A}-1)}\right)\left(l \nu^{2}+\frac{(d_{B}-1)(d_{B}^2+M_{B}^2x_B)}{d_{B}M_{B}(M_{B}-1)}\right)}\\  &\approx  0.000891656p-0.0007866,
\end{align}
which implies that $\rho_p$ is entangled when $0.8821790 \leq p \leq 1$. Fig.\ref{fig:1} shows that our result outperforms that of Ref.\cite{Lu2025}.
\begin{figure}
\includegraphics[width=0.80\textwidth]{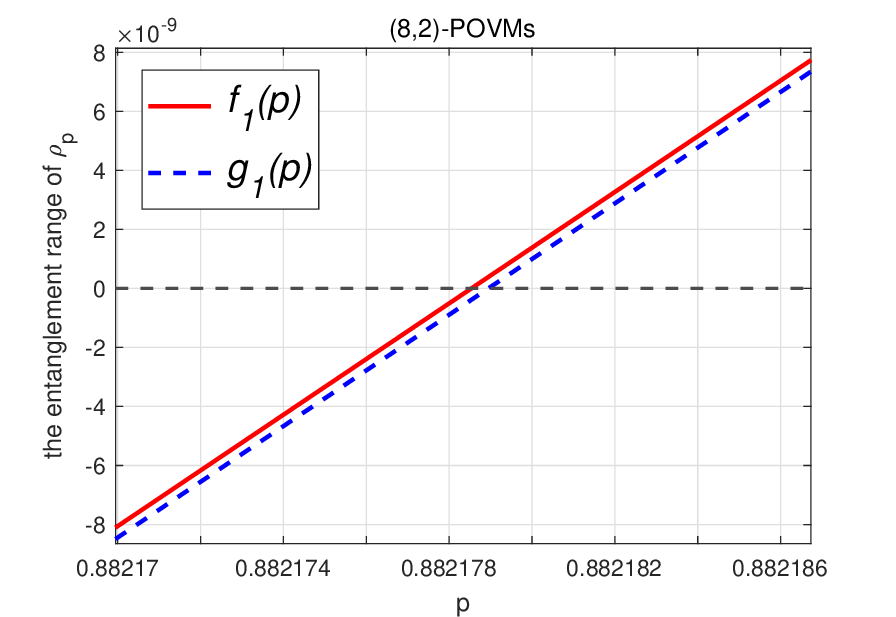}
        \label{fig:1}
\caption{ $f_1(p)$ derived from Theorem \ref{th:1} (solid red line).~\(g_1(p)\)~from the Theorem $1$ of Ref.\cite{Lu2025} (dashed blue line).}
\end{figure}

Example \ref{Ex:1} demonstrates that our method can detect a broader range of entangled states compared to the approach in Ref.\cite{Lu2025}. Here, the values of $a$ and $b$ are so chosen to illustrate the performance of the proposed entanglement criterion under parameter variations. By numerical demonstrations, our criterion is capable of identifying a wider range of entangled states than that of Ref.\cite{Lu2025}. To verify this conclusion, we performed explicit calculations for GSIC-POVMs and MUMs in Appendix \ref{app:A}. The results show that our criterion outperforms that of Ref.\cite{Lu2025}.
\end{example}

\begin{example}\label{Ex:2}
Consider the isotropic states,
\begin{eqnarray}\label{Eq:35}
\rho_{\mathrm{iso}}=q\mid\Psi^{+}\rangle\langle\Psi^{+}|+(1-q)\frac{\mathbb{I}}{d^2},\ \ \  0\leq q\leq1,
\end{eqnarray}
where $\mid\Psi^{+}\rangle=\displaystyle\frac{1}{\sqrt{d}}\sum\limits_{i=0}\limits^{d-1}\mid i\rangle\mid i\rangle$. For $d=3$,  we set $a = (0.1, 0.1)$, $b = (0.05, 0.051)$ and $t = 0.01$  in Theorem \ref{th:1}, Corollary \ref{co:1} and Corollary \ref{co:2}. This leads to the following conclusions:
for the (8,2)-POVMs, Theorem 1 gives that
\begin{align}\label{Eq:36}
f_4(q)\nonumber&\equiv\left\|\mathcal{Q}_{a, b} \left(\rho_{\mathrm{iso}}\right)\right\|_{\mathrm{tr}} - \sqrt{\left(|a|^2+\frac{(d_{A}-1)(d_{A}^2+M_{A}^2x_A)}{d_{A}M_{A}(M_{A}-1)}\right)\left(|b|^2+\frac{(d_{B}-1)(d_{B}^2+M_{B}^2x_B)}{d_{B}M_{B}(M_{B}-1)}\right)} \\&\approx 0.003108 q -0.0007771.
\end{align}

For  $N=1,\, M=9$, the $(N,M)$-POVMs reduce to GSIC-POVMs. By using (\ref{Eq:24}), we obtain
\begin{eqnarray}\label{Eq:37}
f_5(q)\equiv\left\|\mathcal{Q}_{a, b} \left(\rho_{\mathrm{iso}}\right)\right\|_{\mathrm{tr}} -\sqrt{\left(|a|^2+\frac{a_{A}d_{A}^{2}+1}{d_{A}(d_{A}+1)}\right)\left(|b|^2+\frac{a_{B}d_{B}^{2}+1}{d_{B}(d_{B}+1)}\right)} \approx  0.0384 q -0.009616 .
\end{eqnarray}

For  $N=4,\, M=3$, the $(N,M)$-POVMs reduce to MUMs. By using (\ref{Eq:26}), we obtain
\begin{eqnarray}\label{Eq:38}
f_6(q) \equiv \left\|\mathcal{Q}_{a, b} \left(\rho_{\mathrm{iso}}\right)\right\|_{\mathrm{tr}} - \sqrt{\left(|a|^2 + 1 + \kappa_A\right)\left(|b|^2+ 1 + \kappa_B\right)} \approx0.005971 q -0.001493.
\end{eqnarray}

From the conditions \(f_4(q)>0\), \(f_5(q)>0\) and \(f_6(q)>0\), we obtain that \(\rho_{\mathrm{iso}}\) is entangled for \(\frac{1}{4} < q \leq 1\). This result is fully consistent with the known necessary and sufficient condition for the entanglement of isotropic states, namely \(q > \frac{1}{d+1}\) \cite{bertlmann2005optimal}. Therefore, our criterion provides a necessary and sufficient condition for the separability of isotropic states, as shown in Fig.\ref{fig:2}.
\begin{figure}[htp]
\includegraphics[width=0.80\textwidth]{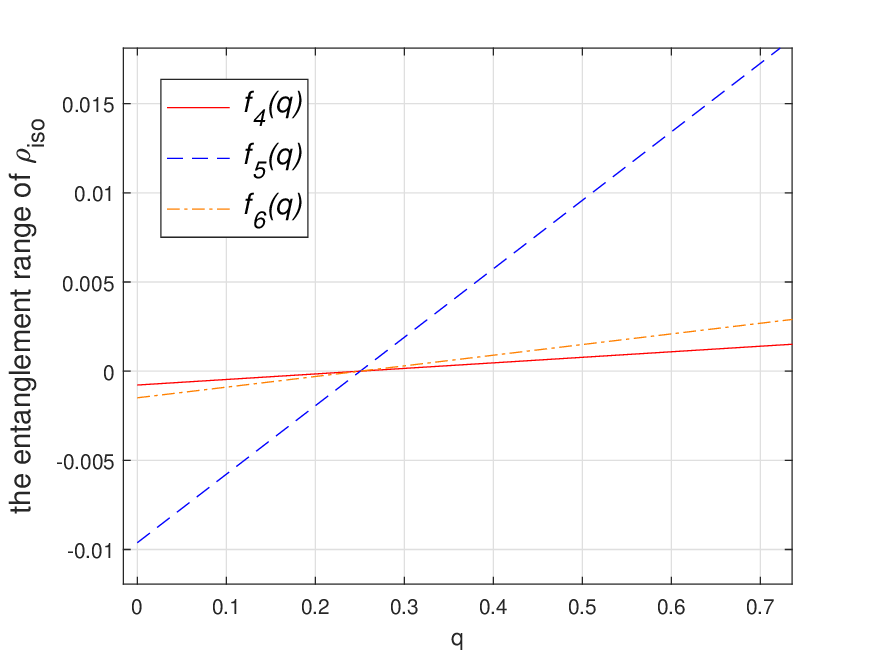}
\caption{~\(f_4(q)\)~in Theorem \ref{th:1} (solid red line),~\(f_5(q)\)~in Corollary \ref{co:1} (dashed blue line), and~\(f_6(q)\)~in Corollary \ref{co:2} (dash-dotted orange line). It can be seen  that~\(\rho_{\mathrm{iso}}\)~is entangled for~\(\frac{1}{4} < q \leq 1\).}
\label{fig:2}
\end{figure}
\end{example}

\begin{example}\label{Ex:3}
Consider the following state $\rho_1(\lambda)$ \cite{bandyopadhyay2005non,halder2019construction},
\begin{align}\label{Eq:39}
\rho_1(\lambda) &= \lambda |\omega_1\rangle \langle \omega_1| + (1 - \lambda)\rho_{BE},
\end{align}
where $\rho_{BE}$ is a bound entangled state given by
\begin{align}\label{Eq:40}
\rho_{BE} &= \frac{1}{4}\left( I_{9} - \sum_{i=1}^{5}|\omega_i\rangle\langle\omega_i|\right),
\end{align}
with
\begin{align}\label{Eq:41}
\nonumber|\omega_1\rangle& = |2\rangle \otimes \frac{1}{\sqrt{2}}(|1\rangle - |2\rangle),\qquad
\nonumber|\omega_2\rangle = |0\rangle \otimes \frac{1}{\sqrt{2}}(|0\rangle - |1\rangle),\\
|\omega_3\rangle &= \frac{1}{\sqrt{2}}(|0\rangle - |1\rangle) \otimes |2\rangle,\qquad
\nonumber|\omega_4\rangle = \frac{1}{\sqrt{2}}(|1\rangle - |2\rangle) \otimes |0\rangle,\\
\left| \omega_5 \right\rangle &= \frac{1}{\sqrt{3}} (|0\rangle + |1\rangle + |2\rangle) \otimes \frac{1}{\sqrt{3}} (|0\rangle + |1\rangle + |2\rangle).
\end{align}
Since \( \rho_1(\lambda) \) has rank $5$, we used Theorem $1$ from Ref.\cite{tang2023enhancing} to compute the range of \(\lambda\). Then we choose \( a = (0.05, 0.06)^T \), \( b = (0.05, 0.05)^T \) and \( t = 0.01 \). In Ref.\cite{Lu2025}, we chose \(\mu = 0.05\), \(\nu = 0.05\) and \(l = 2\). The comparison in Table I demonstrates that our criterion detects entanglement more effectively than the methods given in  Refs.\cite{tang2023enhancing, Lu2025}.
\begin{center}
	\begin{table*}[htp!]
\caption{Entanglement of the state $\rho_1(\lambda)$ in Example \ref{Ex:3}}
		\label{tab:3}
		\begin{tabular}{cccc}
			\hline\noalign{\smallskip}
		 \quad&Results in Ref.\cite{tang2023enhancing}\quad& Results in Ref.\cite{Lu2025}\quad&Our results\\
		\noalign{\smallskip}\hline\noalign{\smallskip}
		
(8,2) - POVMs \quad&$0 \leq \lambda \leq 0.069163655$ \quad &$0 \leq \lambda \leq 0.069163656$ \quad& $0 \leq \lambda \leq  0.069163795$ \\		
		\hspace{1mm}\\
(1,9) - POVMs  \quad&$0 \leq \lambda \leq 0.068848113$ \quad& $0 \leq \lambda \leq 0.068861530$ \quad& $0 \leq \lambda \leq  0.068862672 $ \\
\hspace{1mm}\\
(4,3) - POVMs  \quad&$0 \leq \lambda \leq 0.069160215$ \quad& $0 \leq \lambda \leq 0.069160231$ \quad&  $0 \leq \lambda \leq  0.069160620$ \\
		\noalign{\smallskip}\hline
		\end{tabular}
	\end{table*}
\end{center}
\end{example}

Our entanglement criterion admits a natural generalization to $d$-dimensional multipartite systems.
Let $\rho_{A_1 A_2 \cdots A_n} \in H_{A_1} \otimes H_{A_2} \otimes \cdots \otimes H_{A_n}$ be an $n$-partite state. 
Denote $\{E_{\alpha,k}\}_{\alpha = 1}^{N_{A_q}}$ the $(N,M)$-POVMs on $H_{A_q}$ with efficiency parameters $x_{A_q}$, $q=1,2,...,n$. Denote $\|\mathcal{Q}_{a, b}(\rho_{A_1 \cdots \bar{A}_q \cdots A_n})\|_{\rm tr}$ the matrix given by (\ref{Eq:21}) under bipartition $A_q$ and the rest subsystems. We have the following theorem.

\begin{theorem}\label{th:2}
 If $\rho_{A_1 A_2 \cdots A_n}\in H_{A_1} \otimes H_{A_2} \otimes \cdots \otimes H_{A_n}$ is separable, then
\begin{eqnarray}\label{Eq:42}
\|\mathcal{Q}_{a, b}(\rho_{A_1 \cdots \bar{A}_q \cdots A_n})\|_{\rm tr} &\leq \sqrt{\left(|a|^2+\frac{(d_{A_q}-1)(d_{A_q}^2+M_{A_q}^2x_{A_q})}{d_{A_q}
M_{A_q}(M_{A_q}-1)}\right)\left(|b|^2+\prod_{i=1,i \neq q }^{{n}}\frac{d_{A_i} - 1}{d_{A_i}} \frac{d_{A_i}^2 + M_{A_i}^2 x_{A_i}}{M_{A_i} (M_{A_i} - 1)}\right)}.
\end{eqnarray}
\end{theorem}

\begin{proof}

We first prove that for any separable state $\rho_{A_1 A_2 \cdots A_n} \in H_{A_1} \otimes H_{A_2} \otimes \cdots \otimes H_{A_n}$, we have

\begin{align}\label{Eq:43}
 &\nonumber\sum_{\alpha_1 =1}^{N_{A_1}}\sum_{\alpha_2 =1}^{N_{A_2}}\cdots\sum_{\alpha_n =1}^{N_{A_n}} ~\sum_{k_1 =1}^{M_{A_1}} \sum_{k_2 =1}^{M_{A_2}}\cdots\sum_{k_n =1}^{M_{A_n}} \mathrm{tr}(E_{\alpha_1,k_1}\otimes E_{\alpha_2,k_2}\otimes\cdots\otimes E_{\alpha_n,k_n} \rho_{A_1 A_2 \cdots A_n})^2\\ &\nonumber= \sum_{\alpha_1 =1}^{N_{A_1}}\sum_{\alpha_2 =1}^{N_{A_2}}\cdots\sum_{\alpha_n =1}^{N_{A_n}} ~\sum_{k_1 =1}^{M_{A_1}} \sum_{k_2 =1}^{M_{A_2}}\cdots\sum_{k_n =1}^{M_{A_n}} \mathrm{tr}(E_{\alpha_1,k_1}\rho_{A_1})^2 ~\mathrm{tr}(E_{\alpha_2,k_2}\rho_{A_2})^2~\cdots~\mathrm{tr}(E_{\alpha_n,k_n}\rho_{A_n})^2\\ &\nonumber= \sum_{\alpha_1 =1}^{N_{A_1}} \sum_{k_1 = 1}^{M_{A_1}} \mathrm{tr}(E_{\alpha_1,k_1}\rho_{A_1})^2 \sum_{\alpha_2 =1}^{N_{A_2}} \sum_{k_2 = 1}^{M_{A_2}} \mathrm{tr}(E_{\alpha_2,k_2}\rho_{A_2})^2 \cdots \sum_{\alpha_n =1}^{N_{A_n}} \sum_{k_n = 1}^{M_{A_n}} \mathrm{tr}(E_{\alpha_n,k_n}\rho_{A_n})^2\\ &\leq \prod_{i=1}^{{n}}\frac{d_{A_i} - 1}{d_{A_i}} \frac{d_{A_i}^2 + M_{A_i}^2 x_{A_i}}{M_{A_i} (M_{A_i} - 1)}.
\end{align}
Therefore, we have
\begin{align}\label{Eq:44}
  \left\|\mathscr{Q}_{a,b}(\rho_{A_1 \cdots \bar{A}_q \cdots A_n})\right\|_{\mathrm{tr}} \nonumber
  &=\left\|\sum\limits_{i}p_{i}\mathscr{Q}_{a,b} \left(\rho_i^{{A}_q} \otimes \rho_i^{A_1 \cdots \bar{A}_q \cdots A_n}\right)\right\|_{\mathrm{tr}}\\
  &\leq \sum\limits_{i}p_{i}\left\|\mathscr{Q}_{a,b} \left(\rho_i^{{A}_q} \otimes \rho_i^{A_1 \cdots \bar{A}_q \cdots A_n} \right)\right\|_{\mathrm{tr}}\nonumber\\
  &=\left\|\left(\begin{array}{cc}ab^T & a \sigma_{i_{A_1 \cdots \bar{A}_q \cdots A_n}}^T \\ \tau_{i_{{A}_q}} b^T  \nonumber& \tau_{i_{{A}_q}} \sigma_{i_{A_1 \cdots \bar{A}_q \cdots A_n}}^\mathrm{T}\end{array}\right)\right\|_{\mathrm{tr}} \\
\nonumber&=\left\|\left(\begin{array}{cc}a & \\ \tau_{i_{{A}_q}}  & \end{array}\right)\left(\begin{array}{cc}b^T & \sigma_{i_{A_1 \cdots \bar{A}_q \cdots A_n}}^\mathrm{T}\end{array}\right)\right\|_{\mathrm{tr}} \\
&=\left\|\left(\begin{array}{cc}a & \\ \tau_{i_{{A}_q}}   & \end{array}\right)\right\|_{\mathrm{tr}} \left\|\left(\begin{array}{cc}b^T & \sigma_{i_{A_1 \cdots \bar{A}_q \cdots A_n}}^\mathrm{T}\end{array}\right)\right\|_{\mathrm{tr}} \\
\nonumber&=\left\|\left(\begin{array}{cc}a & \\ \tau_{i_{{A}_q}}   & \end{array}\right)\right\|_{\mathrm{tr}} \left\|\left(\begin{array}{cc}b^T & \sigma_{i_{A_1 \cdots \bar{A}_q \cdots A_n}}^\mathrm{T}\end{array}\right)\right\|_{\mathrm{tr}} \\
\nonumber&\leq \sqrt{\left(|a|^2+\frac{(d_{A_q}-1)(d_{A_q}^2+M_{A_q}^2x_{A_q})}{d_{A_q}
M_{A_q}(M_{A_q}-1)}\right)\left(|b|^2+\prod_{i=1,i \neq q }^{{n}}\frac{d_{A_i} - 1}{d_{A_i}} \frac{d_{A_i}^2 + M_{A_i}^2 x_{A_i}}{M_{A_i} (M_{A_i} - 1)}\right)},
\end{align}
where the last inequality is due to ({\ref{Eq:9}}) and ({\ref{Eq:43}}).
\end{proof}
Violation of the Inequality (\ref{Eq:42}) implies that the multipartite state is entangled. To illustrate our results, let us consider three-qubit systems.  
For \( N=1, d = 2 \) we give the following four matrices \cite{gour2014construction} for any nonzero \( t \in [-\frac{1}{6\sqrt{6}}, \frac{1}{6\sqrt{6}}] \),
\[
E_{1,k} = \frac{1}{4}I + t(G_4 - 6G_{k}), \quad k = 1, 2, 3,
\]
\[
E_{1,4} = \frac{1}{4}I + 3tG_4,
\]
where
\begin{align*}
G_1 = \frac{1}{\sqrt{2}} \begin{bmatrix} 0 & 1 \\ 1 & 0 \end{bmatrix},\quad 
G_2 = \frac{1}{\sqrt{2}} \begin{bmatrix} 0 & i \\ -i & 0 \end{bmatrix},\quad
G_3 = \frac{1}{\sqrt{2}} \begin{bmatrix} 1 & 0 \\ 0 & -1 \end{bmatrix},\quad 
G_4 = \frac{1}{\sqrt{2}} \begin{bmatrix} 1 & 1+i \\ 1-i & -1 \end{bmatrix}.
\end{align*}

By calculation it is easy to show that \(\sum_{k=1}^4 E_{1,k} = I\), \(\mathrm{tr}(E_{1,k})^2 = \frac{1}{8} + 27t^2\) and \(\mathrm{tr}(E_{1,k_1}E_{1,k_2}) = \frac{1}{8} - 9t^2\), \(k_1 \neq k_2 \in \{1, 2, 3, 4\}\). For any nonzero \(t \in [-\frac{1}{6\sqrt{6}}, \frac{1}{6\sqrt{6}}]\), we have
\[
\frac{1}{8} < \mathrm{tr}(E_{1,k})^2 \leq \frac{1}{4}.
\]
Thus \(\{E_{1,k}\}_{k=1}^4\) constitutes a symmetric informationally complete measurement.

\begin{example}\label{Ex:4}
Consider the quantum state,
\[
\rho = \frac{x}{8} I_8 + (1 - x) |\text{GHZ}\rangle\langle\text{GHZ}|, \quad 0 \leq x \leq 1,
\]
where \(|\text{GHZ}\rangle = \frac{1}{\sqrt{2}}(|000\rangle + |111\rangle)\). When $d=2$, from Theorem \ref{th:2}, we have that
$\rho$  is entangled (not fully separable) when the following three inequalities hold:
\begin{eqnarray*}
\|\mathcal{Q}_{a, b}(\rho_{\bar{A}BC})\|_{\text{tr}} > \sqrt{|a|^{2}+\frac{a_{A}d_{A}^{2}+1}{d_{A}(d_{A}+1)}}\sqrt{|b|^{2}+\frac{a_{B}d_{B}^{2}+1}{d_{B}(d_{B}+1)}\frac{a_{C}d_{C}^{2}+1}{d_{C}(d_{C}+1)}},
\end{eqnarray*}
\begin{eqnarray*}
\|\mathcal{Q}_{a, b}(\rho_{A\bar{B}C})\|_{\text{tr}} > \sqrt{|a|^{2}+\frac{a_{B}d_{B}^{2}+1}{d_{B}(d_{B}+1)}}\sqrt{|b|^{2}+\frac{a_{A}d_{A}^{2}+1}{d_{A}(d_{A}+1)}\frac{a_{C}d_{C}^{2}+1}{d_{C}(d_{C}+1)}},
\end{eqnarray*}
\begin{eqnarray*}
\|\mathcal{Q}_{a, b}(\rho_{AB\bar{C}})\|_{\text{tr}} > \sqrt{|a|^{2}+\frac{a_{C}d_{C}^{2}+1}{d_{C}(d_{C}+1)}}\sqrt{|b|^{2}+\frac{a_{A}d_{A}^{2}+1}{d_{A}(d_{A}+1)}\frac{a_{B}d_{B}^{2}+1}{d_{B}(d_{B}+1)}}.
\end{eqnarray*}
For $a = (2, 2)^T$, $b =(2, 0)^T$, and $t = 0.05$, we find that $\rho$ is an entangled state when $0 \leq x < 0.44369$. According to Theorem 3 in Ref.\cite{PhysRevE.107.054134}, $\rho$ is not fully separable if $f(x) = \|G\|_{\text{tr}} - \sqrt{\frac{4a_A+1}{6} \sqrt{\frac{4a_B+1}{6} \sqrt{\frac{4a_C+1}{6}}}} > 0$, which corresponds to \(0 \le x < 0.4130\) for \(t = 0.05\). 
Notably, our results also outperform the range \(0 \le x < 0.2\) reported in Ref.\cite{gao2011separability}.
\end{example}

\section{Conclusions and Discussions}\label{sec:5}

We have proposed a parameterized separability criterion based on \((N,M)\)-POVMs. By introducing vectors \(a\) and \(b\), we constructed an extended correlation matrix \(\mathcal{Q}_{a,b}(\rho_{AB})\), which unifies and generalizes the existing entanglement detection methods based on MUMs and GSIC-POVMs. Examples demonstrate that this criterion outperforms the results in Refs.\cite{Lu2025} and \cite{PhysRevE.107.054134}. When the vectors \(a\) and \(b\) degenerate to zero vectors, our criterion naturally reduces to the result in Ref.\cite{tang2023enhancing}; when \(a\) and \(b\) are vectors of the same dimension and each consists of identical entries, it reduces to the criterion in Ref.\cite{Lu2025}. Furthermore, we have extended this criterion to arbitrary multipartite systems, providing a new tool for multipartite entanglement detection.

Recently, K. Siudzi{\'n}ska introduced GEAMs constructed from generalized equiangular tight frames in Ref.\cite{siudzinska2024informationally}. GEAMs form a broader class of informationally overcomplete POVMs, with \((N,M)\)-POVMs as special cases. Since \((N,M)\)-POVMs and MUMs are themselves overcomplete in many nontrivial cases, informational overcompleteness does not hinder their application. Moreover, GEAMs can be constructed in arbitrary finite dimensions using Hermitian operator bases, and the additional parameters \(b_\alpha\) and \(\gamma_\alpha\) do not introduce essential computational difficulty. For instance, choosing two distinct values of \(b_\alpha\) with all \(\gamma_\alpha=1/N\) gives a simple GEAM outside the \((N,M)\)-POVM framework.
For conical two-design subclasses of GEAMs, the index of coincidence admits an analytical upper bound. Hence, the block-matrix construction developed in this paper can be naturally adapted to GEAMs by replacing the coincidence-index bound for \((N,M)\)-POVMs with the corresponding GEAM bound. This suggests that our approach may be useful for deriving enhanced separability criteria based on more general informationally overcomplete measurements. Moreover, following the method in Ref.\cite{qi2025k}, our approach may also be extended to the study of \(k\)-partite and genuine multipartite entanglement.

    \bigskip
    \noindent{\bf Acknowledgements}
This work is supported by the National Natural Science Foundation of China (NSFC) under Grant No. 12171044; the specific research fund of the Innovation Platform for Academicians of Hainan Province.

\begin{appendices}
\section{Supplement to Example 1}\label{app:A}
When $N = 1$ and $M = 9$, Theorem \ref{th:1} reduces to Corollary \ref{co:1}. In this case, the Hermitian basis operators $G_{\alpha, k}$ are given by the corresponding Gell-Mann matrices(see \cite{gour2014construction} for  details):
\begin{align*}
&G_{11}=\left(
\begin{array}{ccc}
 \frac{1}{\sqrt{2}} & 0 & 0 \\
 0 & -\frac{1}{\sqrt{2}} & 0 \\
 0 & 0 & 0 \\
\end{array}
\right),~~
G_{12}=\left(
\begin{array}{ccc}
 0 & \frac{1}{\sqrt{2}} & 0 \\
 \frac{1}{\sqrt{2}} & 0 & 0 \\
 0 & 0 & 0 \\
\end{array}
\right),~~
G_{13}=\left(
\begin{array}{ccc}
 0 & 0 & \frac{1}{\sqrt{2}} \\
 0 & 0 & 0 \\
 \frac{1}{\sqrt{2}} & 0 & 0 \\
\end{array}
\right),~~
G_{14}=\left(
\begin{array}{ccc}
 0 & -\frac{i}{\sqrt{2}} & 0 \\
 \frac{i}{\sqrt{2}} & 0 & 0 \\
 0 & 0 & 0 \\
\end{array}
\right),~~\\
&G_{15}=\left(
\begin{array}{ccc}
 \frac{1}{\sqrt{6}} & 0 & 0 \\
 0 & \frac{1}{\sqrt{6}} & 0 \\
 0 & 0 & -\sqrt{\frac{2}{3}} \\
\end{array}
\right),~~
G_{16}=\left(
\begin{array}{ccc}
 0 & 0 & 0 \\
 0 & 0 & \frac{1}{\sqrt{2}} \\
 0 & \frac{1}{\sqrt{2}} & 0 \\
\end{array}
\right),~~
G_{17}=\left(
\begin{array}{ccc}
 0 & 0 & -\frac{i}{\sqrt{2}} \\
 0 & 0 & 0 \\
 \frac{i}{\sqrt{2}} & 0 & 0 \\
\end{array}
\right),~~
G_{18}=\left(
\begin{array}{ccc}
 0 & 0 & 0 \\
 0 & 0 & -\frac{i}{\sqrt{2}} \\
 0 & \frac{i}{\sqrt{2}} & 0 \\
\end{array}
\right).~~
\end{align*}
Here, the corresponding parameter $a=\frac{1}{27}+128t^{2}$ with $t\in[-0. 012, 0. 012]$.

\begin{figure}
   \includegraphics[width=0.80\textwidth]{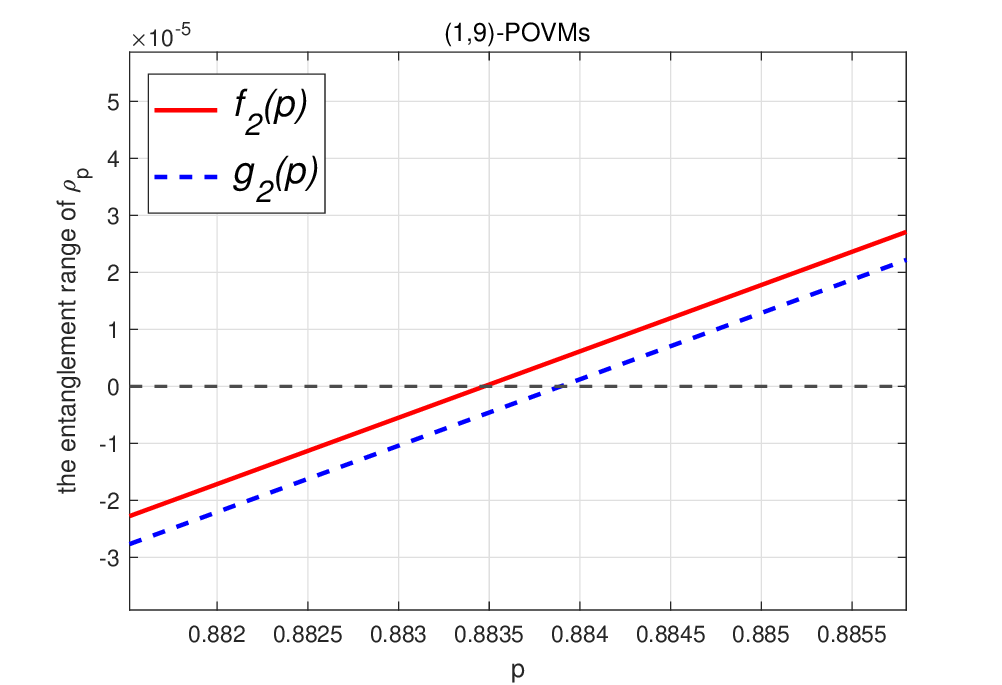}
    \caption{~$f_2(p)$~derived from Corollary \ref{co:1} (solid red line). ~$g_2(p)$~ presented in Corollary $1$ of Ref.\cite{Lu2025} (dashed blue line).}
    \label{fig:3}
\end{figure}

In particular, for $a = (0.1, 0.1)^T$, $b =(0.05, 0.07)^T$ and $t = 0.01$, from Corollary \ref{co:1} we obtain
\begin{align}\label{Eq:45}
f_2(p)\nonumber&\equiv\left\|\mathcal{Q}_{a, b} \left(\rho_p\right)\right\|_{\mathrm{tr}} -\sqrt{\left(|a|^{2}+\frac{a_{A}d_{A}^{2}+1}{d_{A}(d_{A}+1)}\right)\left(|b|^2+\frac{a_{B}d_{B}^{2}+1}{d_{B}(d_{B}+1)}\right)} \\&\approx 0.01100698p -0.00972437.
\end{align}
This shows that $\rho_p$ is entangled within the range $ 0.8834726 \leq p \leq 1$.
 From Corollary $1$ in Ref.\cite{Lu2025}, for $\mu = 0.1$, $\upsilon =0.05$, $l = 2$ and $t = 0.01$,  one obtains
 \begin{align}\label{Eq:46}
g_2(p)\nonumber&\equiv\left\|\mathcal{Q}_{\mu, \nu}^l \left(\rho_p\right)\right\|_{\mathrm{tr}} -\sqrt{\left(l \mu^{2}+\frac{a_{A}d_{A}^{2}+1}{d_{A}(d_{A}+1)}\right)\left(l \nu^{2}+\frac{a_{B}d_{B}^{2}+1}{d_{B}(d_{B}+1)}\right)} \\&\approx 0.01100721 p  -0.00972920,
\end{align}
namely, $\rho_p$ is entangled for $0.8838942 \leq p \leq 1$.
Fig.\ref{fig:3} shows that our result is better than those of Ref.\cite{Lu2025}.

\begin{figure}
   \includegraphics[width=0.80\textwidth]{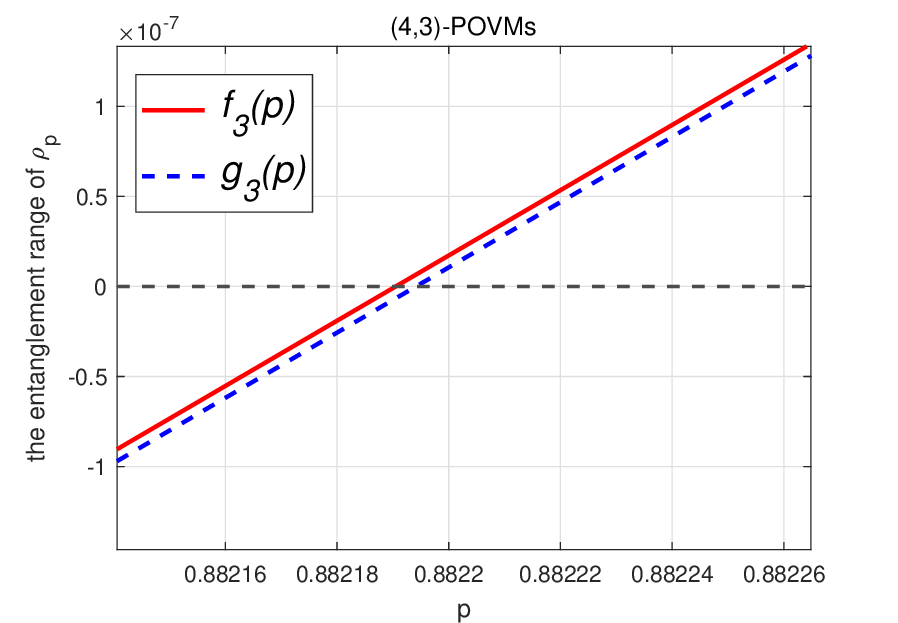}
    \caption{~$f_3(p)$~derived from Corollary \ref{co:2} (solid red line). ~$g_3(p)$~ presented in Remark $2$ of Ref.\cite{Lu2025} (dashed blue line).}
    \label{fig:4}
\end{figure}

When $N = 4$ and $M = 3$, Theorem \ref{th:1} reduces to Corollary \ref{co:2}. In this case, the Hermitian basis operators $G_{\alpha, k}$ are given by the corresponding Gell-Mann matrices (see \cite{kalev2014mutually} for details):
\begin{align*}
G_{11} &= \frac{1}{\sqrt{2}}\left(
\begin{array}{ccc}
0 & 1 & 0 \\
1 & 0 & 0 \\
0 & 0 & 0
\end{array}
\right),~~
G_{12} = \frac{1}{\sqrt{2}}
\left(
\begin{array}{ccc}
0 & -i & 0 \\
i & 0 & 0 \\
0 & 0 & 0
\end{array}
\right),~~
G_{21} = \frac{1}{\sqrt{2}}
\left(
\begin{array}{ccc}
0 & 0 & 1 \\
0 & 0 & 0 \\
1 & 0 & 0
\end{array}
\right),~~
G_{22} = \frac{1}{\sqrt{2}}
\left(
\begin{array}{ccc}
0 & 0 & -i \\
0 & 0 & 0 \\
i & 0 & 0
\end{array}
\right),\\
G_{31} &= \frac{1}{\sqrt{2}}
\left(
\begin{array}{ccc}
0 & 0 & 0 \\
0 & 0 & 1 \\
0 & 1 & 0
\end{array}
\right),~~
G_{32} = \frac{1}{\sqrt{2}}
\left(
\begin{array}{ccc}
0 & 0 & 0 \\
0 & 0 & -i \\
0 & i & 0
\end{array}
\right),~~
G_{41} = \frac{1}{\sqrt{2}}
\left(
\begin{array}{ccc}
1 & 0 & 0 \\
0 & -1 & 0 \\
0 & 0 & 0
\end{array}
\right),~~
G_{42} = \frac{1}{\sqrt{6}}
\left(
\begin{array}{ccc}
1 & 0 & 0 \\
0 & 1 & 0 \\
0 & 0 & -2
\end{array}
\right).
\end{align*}
Here, the parameter $\kappa=\frac{1}{3}+2t^{2}(1+\sqrt{3})^2$ with $t\in[-0.0547,0.3454]$.
In particular, for $a = (0.1, 0.1)^T$, $b =(0.05, 0.07)^T$ and $t = 0.01$ Corollary \ref{co:2} gives rise to
\begin{align}\label{Eq:47}
f_3(p) \nonumber&\equiv \left\|\mathcal{Q}_{a,b} \left(\rho_p\right)\right\|_{\mathrm{tr}} - \sqrt{\left(|a|^2 + 1 + \kappa_A\right)\left(|b|^2+ 1 + \kappa_B\right)} \\&\approx 0.00171208p -0.00151038.
\end{align}
This shows that $\rho_p$ is entangled within the range  $0.8821906 \leq p \leq 1$. From Corollary $2$ in Ref.\cite{Lu2025}, for $\mu = 0.1$, $\upsilon =0.05$, $l = 2$ and $t = 0.01$,  one has
\begin{align}\label{Eq:48}
g_3(p) \nonumber&\equiv \left\|\mathcal{Q}_{\mu, \nu}^l \left(\rho_p\right)\right\|_{\mathrm{tr}} - \sqrt{\left(l\mu^{2} + 1 + \kappa_A\right)\left(l\nu^{2}+ 1 + \kappa_B\right)} \\&\approx 0.00171207p  -0.00151038.
\end{align}
$\rho_p$ is entangled for $0.8821942 \leq p \leq 1$. A comparison with the results in Ref.\cite{Lu2025}, provided in Fig.\ref{fig:4}, shows that our method detects a wider range of entangled states.

\end{appendices}
\end{document}